Best Practices in Statistical Computing

Ricardo Sanchez
Principal Data Scientist, UnitedHealthcare 70127

Beth Ann Griffin
RAND Corporation, Arlington, VA, 22202

Joseph Pane
RAND Corporation, Pittsburgh, PA 15213

Daniel McCaffrey
Educational Testing Service, Princeton, NJ 08541

**Author Footnote:** Ricardo Sanchez is a Principal Data Scientist at UnitedHealthcare (email: ricardo_sanchez@uhc.com); Beth Ann Griffin is Senior Statistician at RAND Corporation Arlington, VA (email: bethg@rand.org); Joseph Pane is a statistical analyst at RAND Corporation Pittsburgh, PA (email: josephp@rand.org ); Daniel McCaffrey is a Principal Research Scientist at Educational Testing Serivces Princeton, NJ (email: dmccaffrey@ets.org).

## Abstract

The world is becoming increasingly complex, both in terms of the rich sources of data we have access to and the statistical and computational methods we can use on data. These factors create an ever-increasing risk for errors in code and the sensitivity of findings to data preparation and the execution of complex statistical and computing methods. The consequences of coding and data mistakes can be substantial. In this paper, we describe the key steps for implementing a code quality assurance (QA) process researchers can follow to improve their coding practices throughout a project to assure the quality of the final data, code, analyses and results. These steps include: (i) adherence to principles for code writing and style that follow best practices, (ii) clear written documentation that describes code, workflow and key analytic decisions; (iii) careful version control, (iv) good data management; and (v) regular testing and review. Following these steps will greatly improve the ability of a study to assure results are accurate and reproducible. The responsibility for code QA falls not only on individual researchers but institutions, journals, and funding agencies as well.

**Keywords: Methodology, Data Management, Version Control**

# 1. Introduction

The world is becoming increasingly complex, both in terms of the rich sources of data we have access to as well as the statistical and computational methods we can use on data. These factors create an ever-increasing risk for errors in our code and sensitivity in our findings to data preparation and execution of complex statistical and computing methods. For example, an analysis of 423 PubMed retraction notices revealed that 18.9% were the result of "analytic errors" related to code and data.[1] In some fields the risk of errors are particularly high. For instance, an analysis on software packages used in fMRI analyses showed that false positive rates could be as high as 70% when null datasets with no relationships were inputted into the software programs.[2] Similarly, approximately one in five papers in genetics research had errors resulting from Microsoft Excel erroneously automatically converting a gene name "Septin 2" to a date "2-Sep".[3] In both cases, the analysts had no safeguards in place to check data accuracy and catch such errors. These problems will only become more frequent and difficult to detect as tools and data analysis workflows become increasingly sophisticated.[4]

The consequences of coding and data mistakes can be substantial. They can require the retraction of published work which can gravely harm the reputations of the authors, their institutions, and the journals publishing retracted articles.[5,3] They can also misinform the public resulting in harmful decisions being made about the potential effectiveness of new treatments.[6,7] Righting those misconceptions can be particularly difficult as has been the case of dissuading people of the link between vaccines and autism (although that misinformation was purposive).[2] They also contribute to the ongoing crisis in the scientific literature where published results fail to replicate.[8-10] The replication crisis in medicine is of particular concern for both patients and providers.[8-10] These consequences erode the basis for the validity of scientific claims and scientific methods as well as the public's trust in science, and consequently, the utility of science.

The increasing complexity of data and the computer code used to process and analyze them suggest these types of problems are not going away. However, programming and data analysis standards that call for clear and complete documentation could help mitigate such problems.[11] For example, knowing exactly what software version was used is useful to ensure results are still valid were a bug found in the future, while knowing which methods were employed is useful for reproducing or replicating a study's findings. Clean and well-documented code can also be verified by reviewers who might be able to catch unintended errors. Moreover, well-documented code, including clear documentation of the analytic workflow and data processing procedures that link back to the source data files, combined with sharing of data and code, can greatly enhance the ability of other analysts to reproduce a study's findings.[12] Sharing of well-documented code also supports replication of the research, because it can provide full details on the exact methods used in the analysis which typically are not included in depth in published reports or journal articles.

Data and code sharing is key to both minimizing coding and analytic mistakes as well as the reproducibility and replication problem.[13] However, this practice must be formally adopted by the community and enforced by the publishers to have an effect. *Nature*, for example, requires that materials, data, code, and protocols be made available. When this is not feasible, any restrictions or exclusions must be fully disclosed.[14] The American Economic Association also requires that data and code are clearly documented and that authors provide non-exclusive access to the data and code.[15]

Sharing code and data, however, is not sufficient. First, secondary users of the code can only catch errors after the fact. A better process is to develop code via a process that avoids errors. Second, code that is not well-developed, including clear documentation, is difficult for other analysts to use and does not make the methods used transparent or easy to replicate in future studies. Code, including the data processing code, needs to be developed following procedures that catch and remove errors in addition to yielding clean and well-documented code that is likely error-free. Moreover, code should be written such that it can be reviewed and used by other analysts to reproduce the study's findings and/or to replicate the methods in future studies. We outline such a process, *Code Quality Assurance* (QA), in this manuscript. The Code QA process is intended to help research teams minimize the likelihood of introducing unintentional errors when working with and processing data. Furthermore, the process is intended to make it easier for reviewers and external research teams to validate and reproduce the results of a given study as well as replicate the study methods in future studies. Other authors also recognize the need for robust code to reduce the risk of error and support reproducibility.[16,17] Taschuk and Wilson (2017) proposed ten simple rules for scientific computing and Vable, Diehl and Glymour (2021) provide a process for reviewing code.[16] Code QA require more than simply following the rules for developing and reviewing code. Code QA is an attitude and process for conducting data analysis that sees quality code and data management procedures as the essential foundation of the conducted research. Code QA begins before the first line of code is written. It is a continuous process, executed through all phases of the research; it is not something that should solely occur at the conclusion of a project. Code QA requires carefully considered and thorough documentation. This means describing the data, explaining the methods, and justifying analytic decisions made through the course of the project. The goal is to provide sufficient detail such that anyone can understand and reproduce the research. Furthermore, the code should be well-written and organized such that it can be easily checked for errors. Code QA should underpin all the work related to code and execution of the statistical methods used in a study.

In this paper, we describe a code QA framework that researchers can use to improve the quality of their programming and data analytic workflows. The ideas are borrowed heavily from concepts in software engineering and were adapted for working with data. Software engineering principles are easier to apply to work that more closely resembles traditional software development, e.g., when creating open-source statistical tools or high-performance simulations. With that in mind, we will focus more on programming for data analysis. What we propose is a general framework for ensuring quality and leave the specifics of the implementation open to interpretation as there is no one-size-fits-all solution to building robust analytic workflows.

The keys of such code QA (see Figure 1) are as follows. First, as noted by Vable, Diehl, and Glymour (2021)[17] it begins at the Principal Investigator (PI) level and requires a team commitment. The commitment begins with the team deciding upon and adhering to principles for code writing and a coding style guide that follows best practices. Next, the team needs to create clear written documentation throughout the lifecycle of a project that describe code, workflow and key analytic decisions. Third, careful version control to track code changes should be employed. Fourth, the team needs good data management processes, and last, the project should test and carefully review code regularly. Following these steps will greatly improve the ability of a study to assure results are accurate and reproducible. As noted, the steps outlined here are not meant to be a one-size-fits-all solution to working with data. Instead this paper is intended to walk through steps that can be applied to help ensure confidence in the output of your analysis.

The steps you choose to implement and how you chose to implement them will depend on many factors, such as the size of the team, size of the data, and complexity of the tools being used.

We thoroughly review each step, in turn, throughout the next sections by way of a case study that can be found here: (https://github.com/jpane24/code-qa). The goal of the original study was to use propensity score weighting methods to estimate the relative effectiveness of two evidence-based treatments on adolescent substance users and assess the potential impact unobserved confounders might have on the analysis. The full study results can be found in Griffin et al (2020).[18] We utilize the case study to showcase an example of how each step of the code QA process can be integrated into an overall practice of data and analysis quality, transparency, and sharing to support accurate and reproducible results.

*1.1 Step 1: Decide upon and adhere to principles for code writing and a coding style guide that follow best practices*

Reducing the risk of errors and producing code that is transparent places requirements on how code is written. To ensure the final code meets those requirements, code writing must follow a set of clearly defined principles and a style guide that adhere to best practices.[16,17] These principles must be established, and the style guide selected before the first line of code is written. Thus, choosing a set of principles for writing code and a coding style guide is the first step in assuring quality in code and results. For code QA, the key is writing code that is readable and well-organized. Principles for writing readable code are different from the principles of software engineering which focus on writing efficient code in terms of performance. Both sets of principles are important; however, code that yields effective computing but is not clear and cannot be understood does not support transparency, reproducibility, or minimize the risk of errors. Code QA should yield code that is both efficient and can be understood by analysts who did not write it.

Style guides are essential to the practice of writing understandable code. They layout rules for naming functions and variables, indenting, and other formatting related issues that make code neater and easier to follow. These rules can sometimes feel arbitrary, the code will likely run whether you follow them or not, but they greatly improve readability. Adherence to a style guide will help collaborators working on the same code base in addition to making it easier for external reviewers to understand the code. Style guides exist for most programming languages (see Table 1) and may be modified to better suit the specific project or team. If a suitable guide cannot be found, the research team should create their own set of rules to follow. The key element is that the rules should be applied consistently throughout the project. Fortunately, this is done easily enough by software tools, or "linters", which will examine a piece of code and flag style guide violations. R has additional add-in packages that can be used to automate the formatting of previously written code to comply with certain style guides. The styler package in R formats code to comply with the tidyverse Style Guide.[19] Figure 2 provides a sample from the code used in our illustrative case study for which all code follows the tidyverse Style Guide.

In addition to the style guides, there are generally accepted best practices for writing "clean" code that primarily focus on code organization.[20] For example, a key element of clean code concerns the utilization of functions. It is generally preferable to use many small functions, less than 100 lines of code, that fully encapsulate discrete and independent functionality rather than one large block of code. There are several benefits to this practice. The first is that it helps the programmer organize and think through the exact purpose of the code. Wrapping code in a function is easier to resuse and makes top-level code, likely stringing together several functions, easier to read. Lastly, it is easier to test and debug code that is broken up into individual components, a topic we will discuss in a later section.

Failing to adhere to styling conventions and best practices contributes to code "smell," a subjective and unquantifiable sense of badness.[21] Although not a problem in and of itself, code smell can be an early warning indicator of poorly written or bad quality code. Elements that contribute to creating code smell include code clutter and poor housekeeping and organization. For example, blocks of duplicated code are highly discouraged as changes to that code can sometimes fail to propagate to each of the copies. This occurs often enough that many modern programming environments will display a warning, much like when Microsoft Word suggests spelling or grammar changes when it finds repeated blocks of code. Along these lines, "ghost" code, i.e., code that is commented out and never executed, should also be avoided. It creates ambiguity by making it difficult for someone reviewing the code to know whether the commented code contributed to the analysis or was used to generate results prior to being commented out. Because of this, dead code should be removed or replaced by a control structure like an if- or switch-statement. Figure 2 provides example of clearly written code purposely meant to avoid a few indicators of code smell discussed above. Well written code is easier to read, develop, maintain, and review. All this works to increase confidence that the code is functioning as intended and contributes to the goal of code QA.

*1.2 Step 2: Create clear written documentation*

The second part of code QA is to create clear written documentation that describes the project, its purpose, goals, and key analytic decisions. It should include a description of input data, generated outputs, and potentially a few executable examples. Although not set in stone, the analytic methods and data workflow should be documented as well. The documentation should contain enough detail about the project that an external reviewer immediately understands the goals and methods used. Documentation also includes a few housekeeping items such as specifying the software license and choosing a method for package and software version management. This ensures the proper setup for the code to run. Documentation can exist as files separate from the code or within the code itself, including both sources, have critical roles in the QA process. There are software tools that help create documentation. For example, these tools create boilerplate comments that function as a starting point for richer documentation. In addition, these tools sometimes understand special formatting can be interpreted and compiled automatically into a user's manual.

An important source of documentation capture is the supporting *README* file (see Figure 3). This file is the first place someone will look before running or reviewing your project. It should give a brief overview, installation instructions, dependencies, directions for running the code, and perhaps a few examples.[16,22] See Figure 3 for an example of a README file (here focused on describing the code-qa github associated with our case study and this paper). It may also be helpful to maintain a document that tracks key analytic decisions along with a description of the intended analytical steps. This file can be stored alongside the *README* file. Lastly, there should be a *LICENSE* file that clearly states which software licensing rules apply to the code. This is especially important for projects that will be released open-source or commercialized. Licensing is a complicated subject and outside the scope of this paper. For more information, the version control and source code management company, GitHub, has created a web portal (https://choosealiense.com/) to help software developers choose between the various license options. Although it is not comprehensive, it is a good entry point into a complex topic. Another site that can be helpful is https://tldrlegal.com/ which helps to try and explain the licenses in easy-to-understand English.

Another critical component of documentation is the version number of all software packages and libraries used in the analysis. This is important as software changes over time and there are no guarantees that the function used today will return the same result in the future. Furthermore, software can change in ways that break code. Common examples include adding a dependency that previously didn't exist or deprecating (i.e., removing) a function. To ensure this does not happen, it is important to document the exact software version number used throughout the project. This could be maintained in a separate file or possibly the README file. For example, the package installer for Python, pip, has a command for exporting a list of every installed pacakage along with their version numbers. This information is typically saved to a text file that can be used when creating new Python virtual environments with the venv module. In R, package information can be displayed using the `sessionInfo()` command or saved using dependency management packages such as Packrat or renv.

The code itself should contain sufficient documentation. Documentation is usually in the form of code comments (see Figure 4 for commented code from our case study). This is English text included in the code files that is not visible to the computer or program running the code. It allows the programmer to add a description of the function and should be helpful to readers trying to understand the code. Style guides offer suggestions for how detailed documentation should be. For example, it is often considered best practice to include a large commented section at the front of a function declaration that briefly explains the purpose of the function and typically includes a description of the expected inputs and outputs. This type of structure is showcased in Figure 4 where the commented text at the beginning describes (1) what the function does (here shows summary statistics for the inputted variables), (2) the needed input variables (here a particular data set and the column name for the variable to be summarized), and (3) the output that will be produced (basic summary statistics and plots of the inputted variable).

In-line comments are text that appear throughout the code that explain, in words, what is being done. Comments are written for people reading the code, which may include the original programmer who could be returning to the project after a long break. Comments should be used sparingly to clarify portions of code that are not obvious when reading the implementation. Good documentation goes beyond just explaining the code functioning but also explains the motivation for the actions being described. When making a key analytic decision, it is important to help someone not only read the code but also document why an analytic decision was made. Figure 5 demonstrates this well for analytic decisions made concerning handling of missing data in the case study showcasing that while checking for missingness patterns in the data, two variables (sncnt and engage) were missing for all observations and therefore should not be included in the imputation. In short, code comments should focus on the why not the how. Comments should not be used to indicate who changed the code or why, that already exists in version control. A final piece of documentation is a detailed explanation of the decisions and inner workings of the code. This is typically done in a journal publication, technical report, or manual.

*1.3 Step 3: Tracking Code Versions*
As the code used for a project develops, it is important to have a mechanism and workflow that can track changes to the code. We do not advocate manual tracking. Rather best practice calls for the use of an automated version tracking system, like Git.[16] Version tracking systems maintain logs of all changes made to the code, creating a record indicating what was deleted or added. They also track additional information such as the date of the change and the name of the person who made the change. All modifications are tracible to the source and can be easily undone. Figure 6 illustrates the use of version control in our case study and an example of multiple users working within the same code repository. This just scratches the surface of what a modern version control system can do. Much can also be said about the best way to track code changes. In fact, articles, blogs and books have been written on the subject.[23,24] Therefore, we will only cover the major points in this paper.

First, versioning systems keep track of every change made, maintaining a historic record of the evolution of the code base. Figure 6 shows an example of four recent documented changes/commits for our illustrative case study. Tracking changes means that any modification to the codebase can be undone by going back to a previous state. Many versioning systems have features that allow one to add comments alongside code changes, making it an effective way to understand the reason for the change (Number 3 in Figure 6). Thus, these tools are especially helpful when working with multiple collaborators who are simultaneously editing the codebase. Individuals will work on their local copy. Collaborators *push* any changes they make to the code for everyone to observe and review. To get the latest changes, each collaborator *pulls* to receive the latest changes and updates made by others. Because collaborators may end up modifying the same section of code, the versioning system also has a way to resolve *conflicts*. Versioning systems are convenient ways to collaborate on the same code. Figure 6 illustrates multiple users interacting with the code (e.g. jpane24 and Sanchez).

Once a versioning system has been chosen, it is highly recommended to develop a workflow. For example, many versioning systems will allow the code to be effectively copied and modified separately from the main body of code. This is convenient for organizing and developing individual features or improvements to the code. It allows those changes to be performed in isolation, without affecting other users, until those changes are ready to be *merged*, or reincorporated, into the main body of code.

To help avoid potential problems and to assist with QA, there are several suggested workflows.

One popular workflow is Git Flow,[25] which proposes that all work, such as bug fixes, new features, and code improvements, be done in separate *branches* or copies of the code until completed and tested. At that point, the code is "merged" or reintegrated with the main branch. The advantage of this approach is that the main branch is in a perpetual "good" state, meaning it only contains code that has been tested and reviewed. This workflow works well for scientific computing, which is why we suggest it. However, there have been some critiques of the technique. For example, long-running feature branches can lead to integration problems and effectively reduce communication by limiting code sharing.[26] One promising alternative is GitHub Flow.[27] Seemingly inspired from Git Flow, it is a less rigorous process of branching and merging that works well in settings like scientific computing where there are typically fewer developers and less emphasis on the ability to rollback or track code version numbers.

Versioning systems track changes to files by doing a line-by-line comparison. As such, they do not handle binary file types like Microsoft Office or PDF documents well. However, good QA includes version control for both the code and its documentation. Hence, documentation should be written if flat text files, using markdown for styling, and its version should be managed alongside the code. Figure 3 and other files with a .md extension in the case study example use markdown for documentation and proper styling of that documentation. Similarly, data should not be put in the versioning systems. This is especially true if the data contains sensitive information and must be protected under the data use agreement. Intermediary data formats and final results can always be regenerated from the code and the original data source files. If data needs to be shared among collaborators, then an appropriate file sharing system should be used for the data.

Versioning systems do a great job of tracking each little change to the code. However, for a more holistic view, major changes and updates to the code should be noted in a text file that is a component of the documentation, called the *NEWS* file. This file tells the users of your code what you changed and how it affects their use of your code. This is a top-level file that exists alongside the README and is typically a first stop for anyone who wants to know what's new or different. This typically occurs when enough changes have been made that you will want to commit a change or increase the version number if you are developing software. Our case study makes use of a NEWS file here: https://github.com/jpane24/code-qa/blob/main/NEWS.md.

*1.4 Step 4: Data Management*

A lot of the focus has been on code; however, proper data management is also an essential component of assuring results are accurate and reproducible. A key element of code QA is the idea that every piece of analysis can be tracked back to its origins in the raw data and that everything can be traced from the input to the code to the output. This can sometimes be tricky if there are multiple processing steps; for example, when data needs to be cleaned and merged or when multiple models are used in series or parallel to generate results. Whatever the workflow, the entire process ultimately depends on knowing exactly what data was used to generate the final results. Hart (2016) does an excellent job laying out ten simple rules for digital data storage.[28] With that in mind, we'd like to emphasize a few key points and fill in a few missing pieces.

Because the input data is an essential element, it is important that it is also well managed. Best practices include storing all raw data in a read-only format or location, preferable in an open and easy to use file format such as CSV, XML or JSON. Data should be treated as immutable. This means the data cannot be modified directly by anyone on the team. Data cleaning, augmentation and transformation is performed on a copy of the data and produces intermediary files (e.g., the .rda files listed in https://github.com/jpane24/code-qa/tree/main/output). As part of the analytic workflow, data preparation is also a part of code QA, meaning it should follow all the practices outlined in this document. Hand editing or using GUI tools to modify data could introduce errors and are not easily reproducible. Therefore, this should be avoided and all changes to the data should be made via scripted code. As with analytic code, code used for data cleaning and preparation should produce the same result every time.[29] It should also be managed following the QA processes described in Steps 1 to 3. When the criteria above cannot be met, for example when working in spreadsheets like Excel, special consideration should be taken to ensure data integrity.[30]

A Digital Object Identifier (DOI) should be obtained for the raw data.[31] This is accomplished by submitting your data to a DOI Registration Agency, typically a third-party provider, that validates the content of the supplied data and metadata. Journal and publications may be able to help in selecting an appropriate repository or may provide guidelines for archiving data. The DOI is a unique, alphanumeric string that can be used to validate that the raw data is the same and has not been modified. In this way, others can be assured that they have access to the exact data used in the research, reducing the risk the results fail to reproduce due to data errors. Of course, special consideration should be given to datasets containing sensitive information, such as those with personally identifiable information (PII). Not all repositories have mechanisms for handling this type of data. Be sure to consult the FAQ or verify with the service before submitting your data.

*1.5 Step 5: Testing and Code Reviews*

Testing is an essential element of code QA. It is composed of two parts. The first part is unit testing in which individual elements of the code are tested. The purpose of a test is to run a function with specific inputs and ensure it returns the correct output for those inputs. Unit tests help specify the code's intended behavior, and are particularly valuable in projects that are developed over many months; it is quite easy to forget about some key detail or corner case that may break the code upon implementing future changes. Writing those cases as unit tests would make this quickly detectable. The goal is to create unit tests for as much of the code as possible. Programming tools can help compute metrics like code coverage which indicates how much of the code is being tested. A good rule of thumb is to create a unit test for each bug, after it is fixed.

The second part is test automation and continuous integration. This means that the unit tests are run regularly, for example each night, on the code. Errors that occur during testing are logged and flagged for the further investigation and possibly repair. Units test should also be run whenever a new change is introduced. For example, if a programmer has recently updated a key function, it is a good idea to run all the unit tests to make sure the change has not inadvertently broken another part of the code. The idea is to continuously monitor the health of the code base.

Finally, a code QA plan should be developed that outlines the policies and procedures for validating the functionality of the code. This includes independent testing of the code to minimize the errors that inherently occur when developing code and reviewing code to ensure it is functioning as described in the analysis plan and supporting documentation. Furthermore, reviews can result in code performance improvements and other optimizations. Thus, a code QA plan should include a provision for code review, where each line of code is seen by at least two eyes.

One way to accomplish this, though often not feasible, is pair programming. In pair programming, two programmers sit side-by-side as they develop the code. Typically, one is at the helm, manning the keyboard, while the other is double checking their work, offering suggestions, or researching solutions. Another way to provide code review is through continuous QA, where code written by one programmer is reviewed by a different programmer prior to integration with the main code base (see Item 4 of Figure 6 for an example in our case study). This ensures a second set of eyes have checked the code and provides a mechanism to continuously monitor the quality of the code base. This should not be confused with continuous integration, which was already discussed. Alternatively, a project can utilize an independent code review by someone outside the team (ideally as early as possible and more than just once) to check for errors and bugs and to ensure the code reproduces the findings from the study.

If one of the end products of the research is a new methodology or tool to implement statistical analyses (e.g., a new R/Python package, SAS/Stata macro, or web/mobile application), additional beta-testing should be done to assess usability concerns. Specifically, a team should conduct a "heuristic evaluation of usability."[32] For heuristic evaluations, a sample of six users is enough to identify at least 85% of the problems associated with the website and newly developed analysis tools.[33] Thus, we recommend new statistical software should undergo three waves of beta testing by six researchers external to the team (two per wave of testing) and from a range of backgrounds. In addition, code developers may wish to provide the beta-testers with a tutorial or instruction on the tool's use along with test data if necessary. However, each beta-tester should be encouraged to use an additional, independent dataset for their review. While vitally important, beta-testing of new software (particularly new R packages) is infrequently done and, as a result, it is not uncommon for developers of R packages to find initial users reporting back many challenges, errors, and bugs when using a new package.

## 2. Conclusion

Increasingly complex data and analyses fuel current research. Data management, preparation and analyses rely on computers which must function appropriately to yield accurate results. Moreover, code documents the analyses and allows for others to reproduce them. Failures of such reproducibility and errors in results have been documented in multiple research areas and can have significant negative impact. As such, there is a clear need for code used in our research and methods development to be fully transparent so our collaborators can easily find errors and external researchers and peers can clearly see the details of the analysis (both in order to reproduce the work and ensure it is error-free as well as to replicate the study methods in future research). These transparency requirements exist across all disciplines, yet adoption of best practices for code transparency is still not nearly as widespread as it should be. There is likely an absence of clear guidance on the best practices for developing code in ways that ensure quality and transparency and insufficient training for researchers and developers who create the code.

This paper provides detailed steps to the code QA process that promote openness (providing others with the data and code used in a scientific analysis) and transparency (requiring the inclusion of additional information, such as detailed explanations of data workflows). We outlined five steps for promoting code QA and use an illustrative case study to highlight sample application of the five steps. Although these steps should be followed throughout the lifecycle of the project, it is important to remember that they each requires some amount of planning. Before any code is written, guides should be agreed upon, documentation should begin, a version system should be selected, data management practices should be outlined, and a testing plan should be developed. By starting early and through consistent application of these steps it is possible to create higher quality code that will be easier to validate (e.g., via peer review) and reproduce.

It is not only the responsibility of research teams to implement code QA. Institutions and journals can greatly help by creating institutional practices and journal guidelines that include detail code QA practice. Funding agencies should also require such QA processes and pay for such work to be part of the project budgets.

Although corrective measures like "journals enforcing standards" and "more time checking notebooks" tend to be viewed as less favorable by researchers, we believe such actions could help improve the quality of research. As noted in the introduction, a notable percentage of research is not following such practices (e.g., almost 20% of the 423 PubMed articles revealed "analytic errors" related to code and data).[34] We imagine the percentage of teams not using code QA is much higher. Making data accessible and providing access to the code could help restore confidence in scientific computing, along with enforcement standards and formalized testing and QA prior to publication. Transparency is key to addressing the biggest problems, like exposing p-hacking, and the smaller issues, like errors in data management or analysis. However, it is not the only tool we should look to for help. Thorough and complete documentation is also necessary. This ensures that data and code will run without error and continue to run the same way each time the analysis is repeated.[12]

We acknowledge that data often cannot be shared publicly due to privacy concerns, which directly stands in the way of reproducibility. One solution is the utilization of synthetic microdata. The R package synthpop: Generating Synthetic Versions of Sensitive Microdata for Statistical Disclosure Control can be used to generate synthetic data to replicate sensitive real values with synthetic values under minimum distortion of the statistical information observed in the true dataset.[35] We demonstrate how to create synthetic dataset in our code repository here: https://github.com/jpane24/code-qa/blob/main/code/synth-creation.R. While the results from a study that cannot directly share its data cannot be reproduced without using the real data, sensitivity analyses can be used on a synthetic dataset, which in turn can be tested by others for reproducibility of the data analysis process. In the future, access to techniques such as homomorphic encryption may allow for validation of computation without needing to decrypt potentially sensitive data first.[36]

Finally, the code QA process we described tacitly assumes all analysis will be conducted, programmed or scripted, i.e., the analysis steps exist as written instructions in text are read and executed by a machine to yield results. The process excludes point-and-click analysis conducted with GUI style tools. Because these types of tools are more difficult to track, we suggest they should be avoided or that more work is needed to enable better tracking of their use from a code QA perspective. We appreciate that GUI tools sometimes are used to make advanced analysis more accessible to a wider audience of non-programmers. In such cases, careful documentation of the data processing and analysis steps (e.g., recording of macros generated and run in the background by the GUI, if possible) will help with QA.

## 3. References

1. Casadevall A, Steen RG, Fang FC. Sources of error in the retracted scientific literature. *FASEB J.* 2014;28(9):3847-3855. doi: 10.1096/fj.14-256735.
2. Eklund A, Nichols TE, Knutsson H. Cluster failure: Why fMRI inferences for spatial extent have inflated false-positive rates. *Proceedings of the National Academy of Sciences.* 2016;113(28):7900-7905. 10.1073/pnas.1602413113.

3. Ziemann M, Eren Y, El-Osta A. Gene name errors are widespread in the scientific literature. *Genome Biology.* 2016;17(1):177. 10.1186/s13059-016-1044-7.
4. Botvinik-Nezer R, Holzmeister F, Camerer CF, et al. Variability in the analysis of a single neuroimaging dataset by many teams. *Nature.* 2020;582(7810):84-88. 10.1038/s41586-020-2314-9.
5. Enserink M. How to avoid the stigma of a retracted paper? Don't call it a retraction. *Science Magazine, American Association for the Advancement of Science.* 2017. doi: 10.1126/science.aan6937.
6. Reinhart CM, Rogoff K. Growth in a Time of Debt. *American Economic Review.* 2010;100(2):573-578.
7. Cassidy J. The Reinhart and Rogoff Controversy: A Summing Up. *The New Yorker*2013.
8. Palus S. Make Research Reproducible. *Sci Am.* 2018;319(4):56-59. doi: 10.1038/scientificamerican1018-56.
9. Estimating the reproducibility of psychological science. *Science.* 2015;349(6251):aac4716. doi: 10.1126/science.aac4716.
10. Begley CG, Ellis LM. Raise standards for preclinical cancer research. *Nature.* 2012;483(7391):531-533. doi: 10.1038/483531a.
11. Vable AM, Diehl SF, Glymour MM. Code Review as a Simple Trick to Enhance Reproducibility, Accelerate Learning, and Improve the Quality of Your Team's Research. *American Journal of Epidemiology.* 2021. 10.1093/aje/kwab092.
12. Chen X, Dallmeier-Tiessen S, Dasler R, et al. Open is not enough. *NatPh.* 2019;15(2):113-119. doi: 10.1038/s41567-018-0342-2.
13. Boulton G. International accord on open data. *Nature.* 2016;530(7590):281-281. doi: 10.1038/530281c.
14. Nature Research. Reporting standards and availability of data, materials, code and protocols. London, U.K.: Springer Nature Limited; 2020.
15. American Economic Association. Data and Code Availability Policy. Nashville, TN: American Economic Association; 2020
16. Taschuk M, Wilson G. Ten simple rules for making research software more robust. *PLOS Computational Biology.* 2017;13(4):e1005412. 10.1371/journal.pcbi.1005412.
17. Vable AM, Diehl SF, Glymour MM. Vable, Diehl, and Glymour respond to "Code Review: An Important Step Towards Reproducible Research". *American Journal of Epidemiology.* 2021. 10.1093/aje/kwab091.
18. Griffin BA, Ayer L, Pane J, et al. Expanding outcomes when considering the relative effectiveness of two evidence-based outpatient treatment programs for adolescents. *J Subst Abuse Treat.* 2020;118:108075. 10.1016/j.jsat.2020.108075.
19. Müller K, Walthert L. Styler: Non-Invasive Pretty Printing of R Code. R package San Francisco, CA GitHub; 2021.
20. Martin RC. *Clean Code: A Handbook of Agile Software Craftsmanship.* London, U.K.: London, U.K.: Pearson Education; 2009
21. Fowler M, Beck K, Brant v, Opdyke W, Roberts D. *Refactoring: Improving the Design of Existing Code.* 1 ed. Boston, MA: Addison-Wesley Professional; 1999
22. Johnson M. Building a Better ReadMe. *Technical Communication.* 1997;44(1):28-36.
23. Loeliger J, McCullough M. *Version Control with Git: Powerful Tools and Techniques for Collaborative Software Development.* Sebastopol, CA: O'Reilly Media, Inc.; 2012

24. Bryan J. Excuse Me, Do You Have a Moment to Talk About Version Control? *The American Statistician.* 2018;72(1):20-27. 10.1080/00031305.2017.1399928.
25. Driessen V. A successful Git branching model. *nvie.com Thoughts and writings by Vincent Driessen.* Vol December 1, 2020. Netherlands2010.
26. Hilton R. A Branching Strategy Simpler than GitFlow: Three-Flow. *Absolutely No Machette Juggling: Ron Hilton's rants about software developments, technology, and sometimes Star Wars*. San Francisco, CA: Zeta Global; 2017.
27. GitHub. Understanding the GitHub flow. 2020; https://guides.github.com/introduction/flow/, December 1, 2020.
28. Hart EM, Barmby P, LeBauer D, et al. Ten Simple Rules for Digital Data Storage. *PLOS Computational Biology.* 2016;12(10):e1005097. 10.1371/journal.pcbi.1005097.
29. If the process involves the use of random numbers, for example when sampling from the data, then it important to see the random number generate and document the seed.
30. Broman KW, Woo KH. Data Organization in Spreadsheets. *The American Statistician.* 2018;72(1):2-10. 10.1080/00031305.2017.1375989.
31. Scientific Data. Recommended Data Repositories. London, U.K. : Springer Nature 2020.
32. Nielsen J. Enhancing the explanatory power of usability heuristics. *ACM Press.* 1994:152--158. 10.1145/191666.191729
33. Nielsen J, Landauer TK. A Mathematical Model of the Findings of Usbility Problems. 1993. https://dl.acm.org/doi/abs/10.1145/169059.169166. Accessed December 1, 2020.
34. Casadevall A, Steen RG, Fang FC. Sources of error in the retracted scientific literature. *The FASEB Journal.* 2014;28(9):3847-3855. https://doi.org/10.1096/fj.14-256735.
35. Nowok B, Raab GM, Dibben C. synthpop: Bespoke Creation of Synthetic Data in R. *2016.* 2016;74(11):26. 10.18637/jss.v074.i11.
36. Paddock S, Abedtash H, Zummo J, Thomas S. Proof-of-concept study: Homomorphically encrypted data can support real-time learning in personalized cancer medicine. *BMC Medical Informatics and Decision Making.* 2019;19(1):255. 10.1186/s12911-019-0983-9.

**Table 1. Style guides for different programming languages**

| Language | Style Guide(s) | Guide URL(s) | Linter(s) | Documentation Tool(s) |
| --- | --- | --- | --- | --- |
| C/C++ | Google C++ Style Guide | https://google.github.io/styleguide/cppguide.html | clang-tidy, cppcheck, cpplint | Doxygen, Sphinx |
| Julia | Style Guide | https://docs.julialang.org/en/v1/manual/style-guide/ | | Documenter |
| Matlab | MATLAB Style Guidelines 2.0 | https://www.mathworks.com/matlabcentral/fileexchange/46056-matlab-style-guidelines-2-0 | mlint | Doxygen, M2HTML, Sphinx |
| Python | PEP 8 | https://www.python.org/dev/peps/pep-0008/ | black, flake8, pylint | PEP 257 |
| R | The tidyverse Style Guide, Google's R Style Guide | https://style.tidyverse.org/, https://google.github.io/styleguide/Rguide.html | lintr, styler | docstring, roxygen2 |
| SAS | | | | Code Diary for SAS, Doxygen |
| Scala | Scala Style Guide | https://docs.scala-lang.org/style/ | scalafmt, scalafix | scaladoc, Sphinx |
| Stata | Suggestions on Stata Programming Style | https://www.stata-journal.com/sjpdf.html?articlenum=pr0018 | | MarkDoc |

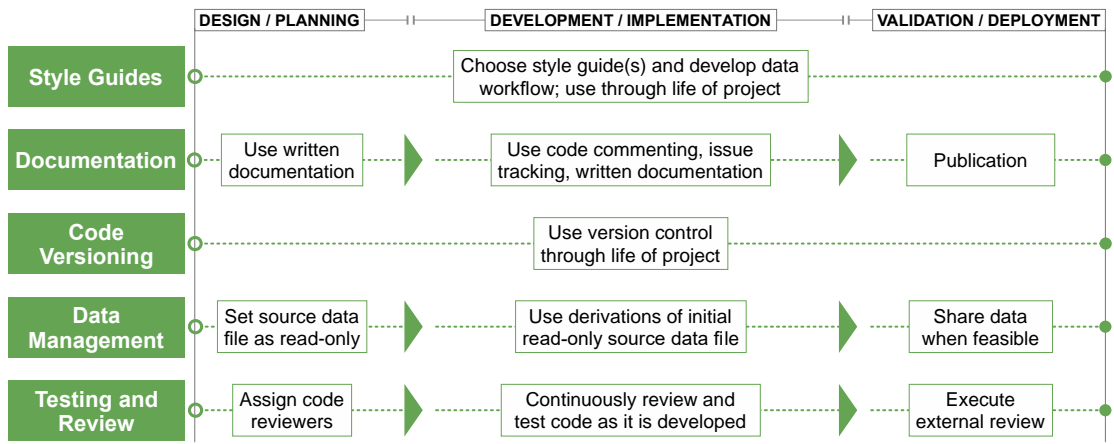

Figure 1. Key strategies for code quality assurance (QA)

```r
30  # Clean data  ==========================================================
31  # Convert any doubles that should be character
32  # Convert character to factor
33  sud <- sud %>%
34    dplyr::mutate(subsgrps_n = case_when(
35      subsgrps_n == 1 ~ "Alc-Marij-Disorder",
36      subsgrps_n == 2 ~ "Other-Drugs",
37      subsgrps_n == 3 ~ "Opioid-Disorder",
38      TRUE ~ NA_character_
39    )) %>%
40    mutate_if(is.character, as.factor)
41
42  # Check for missing values
43  sud %>%
44    dplyr::select_if(., any_NAs) -> var_NAs
45
46  # Count number of missing values in each variable that has >= 1 missing value
47  sud %>%
48    dplyr::select(colnames(var_NAs)) %>%
49    dplyr::summarize_all(sum_NAs)
50
51  # Check for certain missing patterns (within treatment)
52  sud %>%
53    dplyr::select(colnames(var_NAs), treat) %>%
54    dplyr::group_by(treat) %>%
55    dplyr::summarize_all(sum_NAs)
```

Annotations:
1. Descriptive variable names
2. Clear and useful commenting
3. Use of functions fo reduce duplicate code

Figure 2. Sample strategies for writing clean code using case study data

Code snippet can be found here: https://github.com/jpane24/code-qa/blob/main/code/R00-Data-Cleaning.R

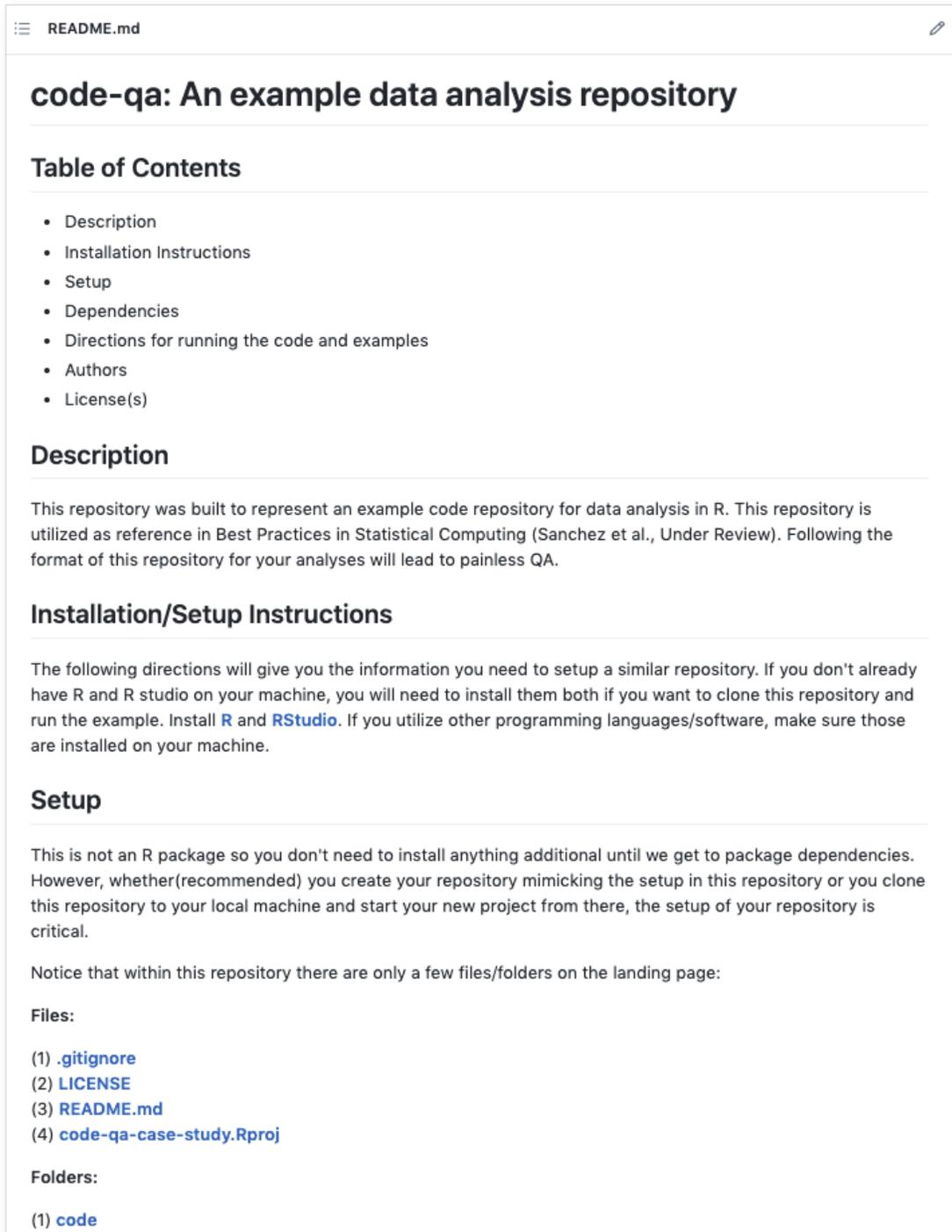

Figure 3. Example README file for the code-qa github

README example from: https://github.com/jpane24/code-qa

Sanchez, Figure 4. TOP```
28    # Sum NAs
29    ds_screener_all <- function(data, var) {
30      # Description: utilize descriptr to show summary stats by variables
31      # Input:
32      # data - a tibble or data.frame
33      # var - a column name (character) in data summary info will be produced for
34      # Output: summary statistics and plots from the descriptr package
35    
36      # We utilize the descriptr package to produce summary statistics
37      # we can change what columns we want to produce summary stats
38      descriptr::ds_summary_stats(
39        data,
40        var
41      )
42    
43      # Frequency table - set bins - produce histogram
44      baseline_outcome <- descriptr::ds_freq_table(data,
45        var,
46        bins = 7
47      )
48      plot(baseline_outcome)
49    }
```

1 Function description
2 Description of inputs and variable types
3 Output description

Figure 4. Example of commented code for a function

Code snippet can be found here: https://github.com/jpane24/code-qa/blob/main/code/helper.R





```r
51    # Check for certain missing patterns (within treatment)
52    sud %>%
53      dplyr::select(colnames(var_NAs), treat) %>%
54      dplyr::group_by(treat) %>%
55      dplyr::summarize_all(sum_NAs)
56    
57    # Note: sncnt and engage are missing in 100% of observations in treatment B.
58    # Analytic decision: Do not impute missing for this analysis...
59    sud <- sud %>%
60      dplyr::select(-colnames(var_NAs))
```

Figure 5. Example of documenting a key analytic decision for handling missing data

Code snippet can be found here: https://github.com/jpane24/code-qa/blob/main/code/R00-Data-Cleaning.R





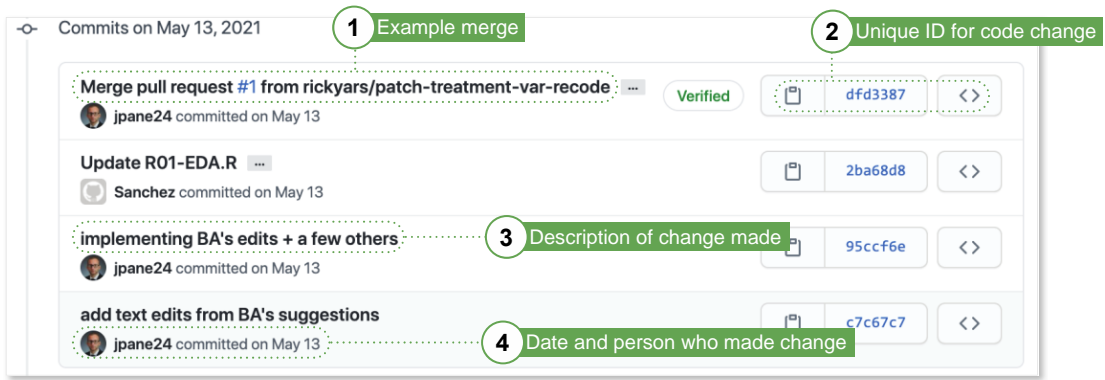

Figure 6. Example of commits and merge